\documentclass{aa}  

\usepackage{graphicx}
\usepackage{txfonts}
\usepackage{xcolor}
\usepackage{hyperref}
\hypersetup{colorlinks, linkcolor=blue, citecolor=blue, urlcolor=blue}
\usepackage{amsmath}
\usepackage{ulem}

\newcommand{\te}{t_{\rm E}}
\newcommand{\thetae}{\theta_{\rm E}}

\newcommand{\pie}{\pi_{\rm E}}

\newcommand{\dl}{D_{\rm L}}

\definecolor{brown}{rgb}{0.59, 0.29, 0.0}
\definecolor{darkgreen}{rgb}{0.0, 0.42, 0.24}
\definecolor{darkblue}{rgb}{0.01, 0.31, 0.59}
\definecolor{darkblue}{rgb}{0.0, 0.25, 0.42}
\definecolor{blue}{rgb}{0.0,0.0,1.0}
\definecolor{green}{rgb}{0.0,1.0,0.0}

\begin{document}

\title{Three faint-source microlensing planets detected via resonant-caustic channel}

\author{
     Cheongho~Han\inst{1} 
\and Andrzej~Udalski\inst{2} 
\and Doeon~Kim\inst{1}
\and Yoon-Hyun~Ryu\inst{3} 
\and Valerio Bozza\inst{4,5}
\\
(Leading authors)\\
     Michael~D.~Albrow\inst{6}   
\and Sun-Ju~Chung\inst{3,7}      
\and Andrew~Gould\inst{8,9}      
\and Kyu-Ha~Hwang\inst{3} 
\and Youn~Kil~Jung\inst{3} 
\and Chung-Uk~Lee\inst{3} 
\and In-Gu~Shin\inst{3} 
\and Yossi~Shvartzvald\inst{10}    
\and Jennifer~C.~Yee\inst{11}      
\and Weicheng~Zang\inst{12}       
\and Sang-Mok~Cha\inst{3,13} 
\and Dong-Jin~Kim\inst{3} 
\and Hyoun-Woo~Kim\inst{3} 
\and Seung-Lee~Kim\inst{3,7} 
\and Dong-Joo~Lee\inst{3} 
\and Yongseok~Lee\inst{3,13} 
\and Byeong-Gon~Park\inst{3,7} 
\and Richard~W.~Pogge\inst{9}
\\
(The KMTNet Collaboration),\\
     Przemek~Mr{\'o}z\inst{2,14} 
\and Micha{\l}~K.~Szyma{\'n}ski\inst{2}
\and Jan~Skowron\inst{2}
\and Rados{\l}aw~Poleski\inst{2} 
\and Igor~Soszy{\'n}ski\inst{2}
\and Pawe{\l}~Pietrukowicz\inst{2}
\and Szymon~Koz{\l}owski\inst{2} 
\and Krzysztof~Ulaczyk\inst{15}
\and Krzysztof~A.~Rybicki\inst{2}
\and Patryk~Iwanek\inst{2}
\and Marcin~Wrona\inst{2}
\and Mariusz~Gromadzki\inst{2}
\\
(The OGLE Collaboration)\\
}

\institute{
     Department of Physics, Chungbuk National University, Cheongju 28644, Republic of Korea  \\ \email{\color{blue} cheongho@astroph.chungbuk.ac.kr}     
\and Astronomical Observatory, University of Warsaw, Al.~Ujazdowskie 4, 00-478 Warszawa, Poland                                                          
\and Korea Astronomy and Space Science Institute, Daejon 34055, Republic of Korea                                                                        
\and Dipartimento di Fisica ``E.~R. Caianiello'', Universit\`a di Salerno, Via Giovanni Paolo II, I-84084 Fisciano (SA), Italy                           
\and Istituto Nazionale di Fisica Nucleare, Sezione di Napoli, Via Cintia, I-80126 Napoli, Italy                                                         
\and University of Canterbury, Department of Physics and Astronomy, Private Bag 4800, Christchurch 8020, New Zealand                                     
\and Korea University of Science and Technology, 217 Gajeong-ro, Yuseong-gu, Daejeon, 34113, Republic of Korea                                           
\and Max Planck Institute for Astronomy, K\"onigstuhl 17, D-69117 Heidelberg, Germany                                                                    
\and Department of Astronomy, The Ohio State University, 140 W. 18th Ave., Columbus, OH 43210, USA                                                       
\and Department of Particle Physics and Astrophysics, Weizmann Institute of Science, Rehovot 76100, Israel                                               
\and Center for Astrophysics $|$ Harvard \& Smithsonian 60 Garden St., Cambridge, MA 02138, USA                                                          
\and Department of Astronomy, Tsinghua University, Beijing 100084, China                                                                                 
\and School of Space Research, Kyung Hee University, Yongin, Kyeonggi 17104, Republic of Korea                                                           
\and Division of Physics, Mathematics, and Astronomy, California Institute of Technology, Pasadena, CA 91125, USA                                        
\and Department of Physics, University of Warwick, Gibbet Hill Road, Coventry, CV4 7AL, UK                                                               
}
\date{Received ; accepted}

\abstract
{}
{
We conducted a project of reinvestigating the 2017--2019 microlensing data collected by the 
high-cadence surveys with the aim of finding planets that were missed due to the deviations of 
planetary signals from the typical form of short-term anomalies.
}
{
The project led us to find three planets including KMT-2017-BLG-2509Lb, OGLE-2017-BLG-1099Lb, 
and OGLE-2019-BLG-0299Lb.  The lensing light curves of the events have a common characteristic 
that the planetary signals were produced by the crossings of faint source stars over the resonant 
caustics formed by giant planets located near the Einstein rings of host stars.
}
{
For all planetary events, the lensing solutions are uniquely determined without any degeneracy. 
It is estimated that the host masses are in the range of $0.45\lesssim M/M_\odot \lesssim 0.59$, 
which corresponds to early M to late K dwarfs, and thus the host stars are less massive than the 
sun.  On the other hand, the planets, with masses in the range of $2.1\lesssim M/M_{\rm J}\lesssim 
6.2$, are heavier than the heaviest planet of the solar system, that is, Jupiter.  The planets in 
all systems lie beyond the snow lines of the hosts, and thus the discovered planetary systems, 
together with many other microlensing planetary systems, support that massive gas-giant planets 
are commonplace around low-mass stars.  We discuss the role of late-time high-resolution imaging 
in clarifying resonant-image lenses with very faint sources.
}
{}


\maketitle

\begin{table*}[htb]
\small
\caption{Source location, alert date, and baseline magnitude\label{table:one}}
\begin{tabular}{llllcc}
\hline\hline
\multicolumn{1}{c}{Event}                  &
\multicolumn{1}{c}{(RA,DEC)}               &
\multicolumn{1}{c}{$(l, b)$}               &
\multicolumn{1}{c}{Alert date}             &
\multicolumn{1}{c}{$I_{\rm base}$}         \\
\hline
KMT-2017-BLG-2509   & (17:42:21.57, -26:19:01.74)     &   $(1^\circ\hskip-2pt .853, 1^\circ\hskip-2pt.987)$   & postseason    & 19.96   \\
\hline              
OGLE-2017-BLG-1099/ & (17:35:51.42, -29:35:09.10)     &   $(-1^\circ\hskip-2pt.679, 1^\circ\hskip-2pt.461)$   & 2017-06-13/   & 20.61   \\
KMT-2017-BLG-2336   &                                 &                                                       & postseason    &         \\
\hline
OGLE-2019-BLG-0299/ & (17:46:43.07,-23:35:03.52)      &   $(4^\circ\hskip-2pt.702, 2^\circ\hskip-2pt.570)$    & 2019-03-16/   &  20.04  \\
KMT-2019-BLG-2735   &                                 &                                                       & postseason    &         \\
\hline
\end{tabular}
\end{table*}

\begin{table*}[htb]
\small
\caption{Data sets, fields, and observational cadences\label{table:two}}
\begin{tabular}{llllcc}
\hline\hline
\multicolumn{1}{c}{Event}           &
\multicolumn{1}{c}{Data sets}       &
\multicolumn{1}{c}{Field}           &
\multicolumn{1}{c}{Cadence}         \\
\hline
KMT-2017-BLG-2509   & KMTA, KMTC, KMTS   &   KMT18      &  1 hr          \\
\hline                                                                
OGLE-2017-BLG-1099  & OGLE,              &   BLG654.2   & 1/3--1 day     \\
                    & KMTA, KMTC, KMTS   &   KMT14      & 1 hr           \\
\hline                                                                                  
OGLE-2019-BLG-0299  & OGLE,              &   BLG632.10  & 1/3--1 day     \\
                    & KMTA, KMTC, KMTS   &   KMT20      & 2.5 hr         \\
\hline
\end{tabular}
\end{table*}

\section{Introduction}\label{sec:one}

During the early phase of planetary microlensing experiments, for example, OGLE \citep{Udalski1994},
MACHO \citep{Alcock1993}, EROS \citep{Aubourg1993} surveys, for which the annual detection rate of 
lensing events was several dozens, individual events could be thoroughly inspected to check the 
existence of planet-induced anomalies in the light curves of the lensing events. With the advent of 
high-cadence surveys, that is, MOA \citep{Sumi2013}, OGLE-IV \citep{Udalski2015}, KMTNet \citep{Kim2016}, 
the detection rate has soared to more than 3000 per year. With the greatly increased number of events 
together with the large quantity of data for each event, planetary signals for some events may escape 
detection.

Two projects have been carried out since 2020 with the aim of finding missed planets in the survey 
data collected in and before the 2019 season.  One project has been conducted to find planet-induced 
anomalies via an automatized algorithm.  \citet{Hwang2021} applied the automated AnomalyFinder software 
\citep{Zang2021} to the 2018--2019 lensing light curves from the $\sim 13~{\rm deg}^2$ of sky covered by
the six KMTNet prime fields with cadences $\geq 2~{\rm hr}^{-1}$. From this investigation, they reported 
6 newly detected planets with mass ratios $q<2\times 10^{-4}$, including OGLE-2019-BLG-1053Lb, 
KMT-2019-BLG-0253Lb, OGLE-2018-BLG-0506Lb, OGLE-2018-BLG-0516Lb, OGLE-2019-BLG-1492Lb, and 
OGLE-2018-BLG-0977. The signals of these planets were not only very short but also weak, and thus 
they had not been noticed from visual inspections.  A more extensive searches for missing planetary 
signals are underway by applying the automated algorithm to all the lensing light curves detected by 
the KMTNet survey (Y.~K.~Jung et al.\ 2021, in preparation).

The other project has been conducted by visually inspecting the previous survey data to find unnoticed 
planetary signals. This project led to the discoveries of 10 planets. The first planet was KMT-2018-BLG-0748Lb 
\citep{Han2020b}, for which the planetary signal had been missed due to the faintness of the source combined 
with relatively large finite-source effects. This discovery was followed by the detections of KMT-2016-BLG-2364Lb, 
KMT-2016-BLG-2397Lb, OGLE-2017-BLG-0604Lb, and OGLE-2017-BLG-1375Lb, for all of which the lensing events involved 
faint source stars \citep{Han2021a}, KMT-2019-BLG-1339Lb, for which the planetary signal was partially covered 
\citep{Han2020a}, KMT-2018-BLG-1976Lb, OGLE-2019-BLG-0954Lb, KMT-2018-BLG-1996Lb, for which the planetary 
signals were produced through a non-caustic-crossing channel, and thus weak \citep{Han2021b}, and 
KMT-2018-BLG-1025Lb, for which the planetary signal had been missed due to the low mass ratio of the planet 
($q\sim 0.8\times  10^{-4}$ or $1.6\times 10^{-4}$ for two degenerate solutions) together with the 
non-caustic-crossing nature of its planetary signal \citep{Han2021c}.  Visually inspecting missing planets 
can provide various types of planetary signals that are prone to being missed, and thus can help to develop 
a more complete algorithm for automatized planet detections.

In this paper, we report three additional planets with similar planetary signatures and event 
characteristics found from the (visual) inspection of the 2017--2019 season data collected by 
the OGLE and KMTNet surveys.  The common feature of these planetary events is that the planetary 
signals were produced by the crossings of faint source stars over the resonant caustics formed by 
giant planets located near the Einstein rings of the lens systems.  Detecting such planetary signals 
requires a visual inspection of lensing light curves, because the durations of the planet-induced 
anomalies comprise important portions of the event durations, making them difficult to be detected 
by an automatized system that is optimized to detecting very short-term anomalies.  An example of 
such a planetary event is found in the case of OGLE-2016-BLG-0596 \citep{Mroz2017}.

For the presentation of the work, we organize the paper as follows. In Sect.~\ref{sec:two}, we 
mention the observations of the lensing events and the acquired data. In Sect.~\ref{sec:three}, 
we mention the characteristics of the anomalies in the lensing light curves of the events, and 
describe the detailed analyses conducted to explain the observed anomalies. In Sect.~\ref{sec:four}, 
we specify the source types of the events, and constrain the angular Einstein radii. In 
Sect.~\ref{sec:five}, we determine the physical parameters of the planetary systems using the 
observables of the events.  In Sect.~\ref{sec:six}, we discuss the role of high-resolution
follow-up observations for faint-source planetary events with resonant caustic features in 
clarifying the nature of the lens system.  A summary of the results and conclusion are presented 
in Sect.~\ref{sec:seven}.

\section{Observations and data}\label{sec:two}

The newly reported three planetary lensing events are KMT-2017-BLG-2509, 
OGLE-2017-BLG-1099/KMT-2017-BLG-2336, and OGLE-2019-BLG-0299/KMT-2019-BLG-2735. The first 
event was detected solely by the KMTNet survey, and the latter two events were detected by 
both the OGLE and KMTNet surveys. Hereafter, we designate the events by the identification 
numbers of the surveys that first found the lensing events. In Table~\ref{table:one}, we 
list the equatorial and galactic coordinates of the source stars, alert dates, and $I$-band 
baseline magnitudes, $I_{\rm base}$, of the individual events. The notation ``postseason'' 
indicates that the event was detected from the postseason inspection of the data.

After one year test observations in the 2015 season, the KMTNet group has carried out a 
lensing survey toward the Galactic bulge field since 2016 with the use of three identical 1.6~m 
telescopes that are distributed in the three continents of the Southern Hemisphere, at the 
Siding Spring Observatory in Australia, the Cerro Tololo Interamerican Observatory in Chile, 
and the South African Astronomical Observatory in South Africa. We designate the individual 
telescopes as KMTA, KMTC, and KMTS, respectively. The camera mounted on each telescope has a 
4~deg$^2$ field of view.  The OGLE team has conducted a lensing survey since its commencement 
of the first phase experiment in 1992 \citep{Udalski1994}, and now the survey is in the fourth 
phase (OGLE-IV) with the use of an upgraded camera yielding a 1.4~deg$^2$ field of view \citep{Udalski2015}. 
The OGLE telescope with an aperture of 1.3~m is located at the Las Campanas Observatory in Chile. 
For both surveys, images of the events were obtained mainly in the $I$ band, and a fraction of 
$V$-band images were acquired for the source color measurements.  In Table~\ref{table:two}, we 
list the data sets used in the analysis, the fields of the individual surveys, and the cadence 
of observations.  We note that the KMTNet cadences of the events, ranging 1.0--2.5~hr, are 
substantially lower than the cadence of the prime fields, 15~min, and this partially contributed 
to the difficulty of identifying the planetary nature of the events.

\begin{table}[t]
\small
\caption{Error adjustment factors\label{table:three}}
\begin{tabular*}{\columnwidth}{@{\extracolsep{\fill}}lcccc}
\hline\hline
\multicolumn{1}{c}{Event}            &
\multicolumn{1}{c}{Data set}            &
\multicolumn{1}{c}{$k$}            &
\multicolumn{1}{c}{$\sigma_{\rm min}$ (mag)}            &
\multicolumn{1}{c}{$N_{\rm data}$}           \\
\hline
KMT-2017-BLG-2509  & KMTA & 1.054   &  0.020   & 200     \\
                   & KMTC & 1.330   &  0.020   & 489     \\
                   & KMTS & 1.192   &  0.020   & 430     \\
\hline
OGLE-2017-BLG-1099 & OGLE & 1.179   &  0.010   & 366     \\
                   & KMTA & 1.149   &  0.020   & 139     \\
                   & KMTC & 1.054   &  0.010   & 183     \\
                   & KMTS & 1.025   &  0.030   & 225     \\
\hline
OGLE-2019-BLG-0299 & OGLE & 1.265   &  0.010   & 249     \\
                   & KMTA & 1.195   &  0.040   & 53      \\
                   & KMTC & 1.545   &  0.020   & 395     \\
                   & KMTS & 1.179   &  0.020   & 354     \\
\hline
\end{tabular*}
\end{table}

\begin{figure}[t]
\includegraphics[width=\columnwidth]{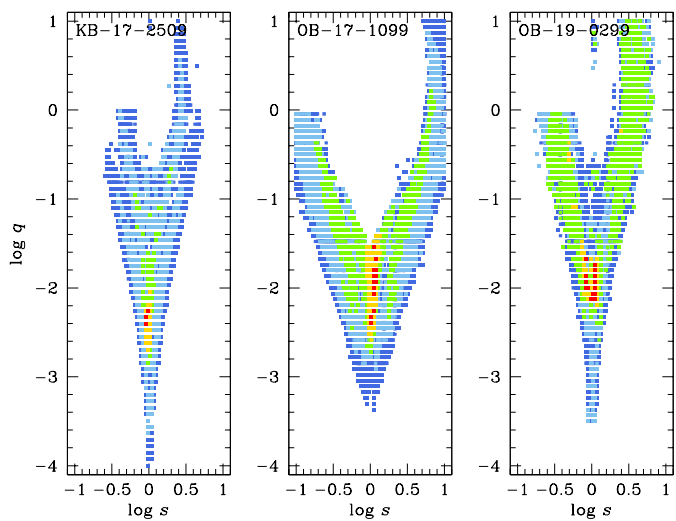}
\caption{
$\Delta\chi^2$ maps on the $\log s$--$\log q$ parameter plane 
obtained from the first-round modeling of the individual events.
Color coding is set to represent points with 
$\Delta\chi^2\leq n(1^2)$ (red),
$\Delta\chi^2\leq n(2^2)$ (yellow),
$\Delta\chi^2\leq n(3^2)$ (green),
$\Delta\chi^2\leq n(4^2)$ (cyan), and 
$\Delta\chi^2\leq n(5^2)$ (blue),
where $n=5$, 3, and 9 for the KMT-2017-BLG-2509, OGLE-2017-BLG-1099, and
OGLE-2019-BLG-0299, respectively.
}
\label{fig:one}
\end{figure}

The data used in the analyses were reduced using the photometry pipelines of the individual
survey groups. These pipelines, developed by \citet{Albrow2009} for KMTNet and by \citet{Wozniak2000} 
for OGLE, utilize the difference imaging technique \citep{Tomaney1996, Alard1998}, which is 
optimized for the photometry of stars in very dense star fields. A subset of the KMTC data were 
additionally processed using the pyDIA photometry code \citep{Albrow2017} to construct color-magnitude 
diagrams (CMD) of stars around the source stars and to specify the source types. See Sect.~\ref{sec:four} 
for details.  For the data used in the analyses, we readjusted the error bars of the photometric data by 
$\sigma = k(\sigma_{\rm min}^2+ \sigma_0^2)^{1/2}$ following the prescription depicted in \citet{Yee2012}.  
Here $\sigma_0$ denotes the error bar from the pipelines, $\sigma_{\rm min}$ is a factor used to make the 
scatter of data consistent with error bars, and $k$ is a scaling factor used to make the $\chi^2$ per 
degree of freedom for each data equal to unity.  In Table~\ref{table:three}, we list the error-bar 
readjustment factors and the number of data, $N_{\rm data}$, of the individual data sets.

\section{Analysis}\label{sec:three}

In this section, we present the detailed analyses of the individual lensing events. The light curves
of all events exhibit anomaly features that include caustic crossings, and thus we model the light
curves under a binary-lens and single-source (2L1S) interpretation.

\begin{figure}[t]
\includegraphics[width=\columnwidth]{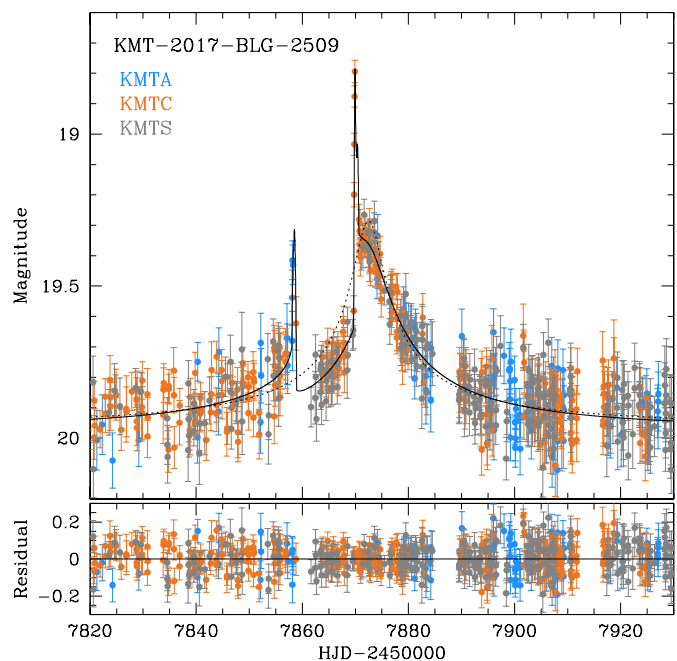}
\caption{
Light curve of KMT-2017-BLG-2509.  The dotted and solid curves plotted over the data points represent 
the 1L1S and 2L1S models, respectively. The residual of the 2L1S model is shown in the lower panel.
The colors of the telescopes in the legend are set to match those the data points.
}
\label{fig:two}
\end{figure}

The modeling is done by searching for the set of the lensing parameters that best explain the observed 
data. The first three of these parameters describe the source--lens approach, including $(t_0, u_0, \te)$, 
which represent the time of the closest source approach to a reference position of the lens, the separation 
(normalized to the angular Einstein radius $\thetae$) between the source and the lens reference position 
at $t_0$, and the event time scale, respectively.  For the reference position of the lens, we use the 
center of mass for a binary lens with a separation less than 
$\thetae$ (close binary), and the effective position, defined by \citet{Stefano1996} and \citet{An2002}, for 
the lens with a separation greater than $\thetae$ (wide binary). The next three parameters $(s, q, \alpha)$ 
describe the lens binarity, and they denote the projected separation (normalized to $\thetae$) and mass ratio 
between the lens components, $M_1$ and $M_2$, and the angle between the source motion and the binary axis 
(source trajectory angle), respectively. The last parameter is $\rho$, which is defined as the ratio  of the 
angular source radius, $\theta_*$, to $\thetae$, that is, $\rho=\theta_*/\thetae$ (normalized source radius), 
and it is included to account for finite-source effects that affect the caustic-crossing parts of lensing light 
curves.  We incorporate limb-darkening effects in the finite magnification computations by modeling the surface 
brightness variation of the source as $S\propto 1-\Gamma(1-3\cos\phi/2)$, where $\Gamma$ is the limb-darkening 
coefficient, $\phi$ represents the angle between two lines extending from the source center: one to the 
observer and the other to the source surface. As will be discussed in Sect.~\ref{sec:four}, the source stars 
of all events have similar spectral types, early K-type main-sequence stars, and thus we adopt 
an $I$-band limb-darkening coefficient of 
$\Gamma_I= 0.5$ 
from \citet{Claret2000}, assuming that the effective temperature, surface gravity, and turbulence velocity 
are $T_{\rm eff}=5000$~K, $\log(g/g_\odot)=0.05$, and $v_{\rm turb}=2~{\rm km}~{\rm s}^{-1}$, respectively.

For a fraction of events with long time scales comprising an important protion of a year, lensing light 
curves may deviate from the form expected from a rectilinear lens-source motion. One cause for this deviation 
is the motion of an observer induced by the orbital motion of Earth \citep[microlens-parallax effect:][]{Gould1992}, 
and the other is the orbital motion of the binary lens \citep[lens-orbital effects:][]{Dominik1998}.  It is 
expected that the signature of the microlens parallax, $\pie$, will be small for OGLE-2017-BLG-1099 and 
OGLE-2019-BLG-0299 due to their short time scales, which are $\sim 19$~days and $\sim 30$~days, respectively, 
but the signature might not be small for KMT-2017-BLG-2509 due to its relatively long time scale of 
$\sim 67$~days.  We check these higher-order effects and find that it is difficult to detect the higher-order 
signatures for any of the events, mainly because the photometric quality is not high enough to detect the subtle 
deviations induced by these higher-order effects.  A microlens parallax can provide an important constraint on 
the physical lens parameters.  As will be discussed in Sect.~\ref{sec:five}, the uncertainties of the physical 
lens parameters are large for all events due to the absence of the $\pie$ constraint.

\begin{figure}[t]
\includegraphics[width=\columnwidth]{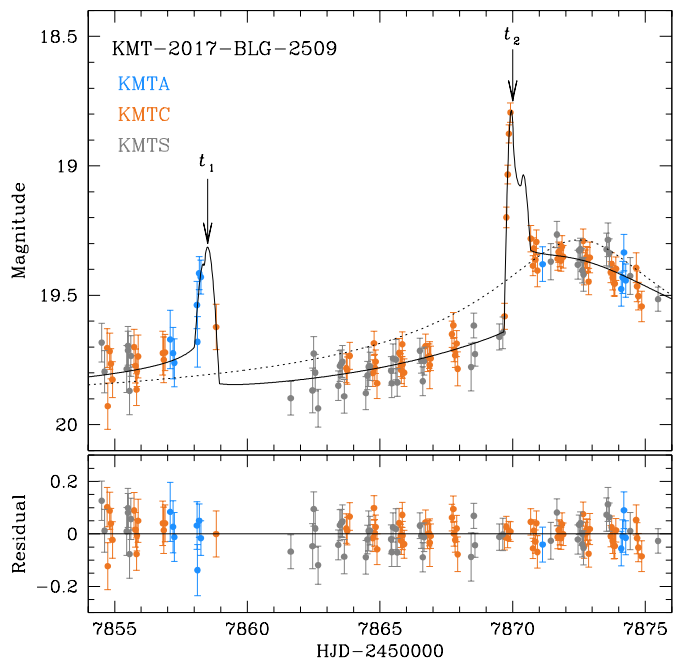}
\caption{
Zoom-in view of the major anomaly region of KMT-2017-BLG-2509. Notations are same as those in Fig.~\ref{fig:two}.
}
\label{fig:three}
\end{figure}

\begin{table*}[t]
\small
\caption{Lensing parameters\label{table:four}}
\begin{tabular}{llllcc}
\hline\hline
\multicolumn{1}{c}{Parameter}      &
\multicolumn{1}{c}{KMT-2017-BLG-2509}      &
\multicolumn{1}{c}{OGLE-2017-BLG-1099}      &
\multicolumn{1}{c}{OGLE-2019-BLG-0299}      \\
\hline
$t_0$ (HJD$^\prime$)       &  $7872.205 \pm 0.158$   &  $7917.336 \pm 0.004$    &  $8560.239 \pm 0.025$   \\
$u_0$                      &  $0.066 \pm  0.007  $   &  $0.004 \pm 0.001   $    &  $0.056 \pm 0.002   $   \\
$\te$ (days)               &  $67.39 \pm  5.37   $   &  $18.87 \pm 1.57    $    &  $29.82 \pm 1.05    $   \\
$s$                        &  $0.925 \pm  0.007  $   &  $1.137 \pm 0.014   $    &  $0.990 \pm 0.002   $   \\
$q$ (10$^{-3}$)            &  $4.366 \pm  0.534  $   &  $6.420 \pm 0.779   $    &  $10.037 \pm 0.637  $   \\
$\alpha$ (rad)             &  $2.531 \pm  0.027  $   &  $0.065 \pm 0.014   $    &  $1.277 \pm 0.0186  $   \\
$\rho$ (10$^{-3}$)         &  $1.927 \pm  0.301  $   &  $1.539 \pm 0.166   $    &  $< 3.5             $   \\
$f_s$                      &  $0.008 \pm  0.001  $   &  $0.012 \pm 0.001   $    &  $0.065 \pm 0.002   $   \\
$f_b$                      &  $0.155 \pm  0.001  $   &  $0.137 \pm 0.001   $    &  $0.018 \pm 0.002   $   \\
\hline
\end{tabular}
\tablefoot{ 
${\rm HJD}^\prime\equiv {\rm HJD}-2450000$.
}
\end{table*}

The 2L1S modeling is carried out in two steps.  In the first step, the binary lensing parameters 
$s$ and $q$ are searched for using a grid approach with different starting values of $\alpha$ 
evenly distributed in the 0 -- $2\pi$ range with 21 grids, while the 5 non-grid lensing parameters, 
that is, $(t_0, u_0, \te ,\alpha, \rho)$, are found using a downhill approach based on the Markov 
Chain Monte Carlo (MCMC) algorithm.  The ranges of the grid parameters are $-1.0\leq \log s < 1.0$ 
and $-4.0\leq \log q < 1.0$, and they are divided into 70 grids.  We then identify local solutions 
that appear in the $\Delta\chi^2$ map on the $s$--$q$ parameter plane. In the second step, we refine 
the individual local solutions found in the first step by letting all parameters vary.  Then the global 
solution is found by comparing $\chi^2$ values of the local solutions, if there is more than one.  
Figure~\ref{fig:one} shows the $\Delta\chi^2$ maps on the $\log s$--$\log q$ parameter plane obtained 
from the first-step modeling of the individual events.  It shows that there exist a single local 
solution for all events.

\begin{figure}[t]
\includegraphics[width=\columnwidth]{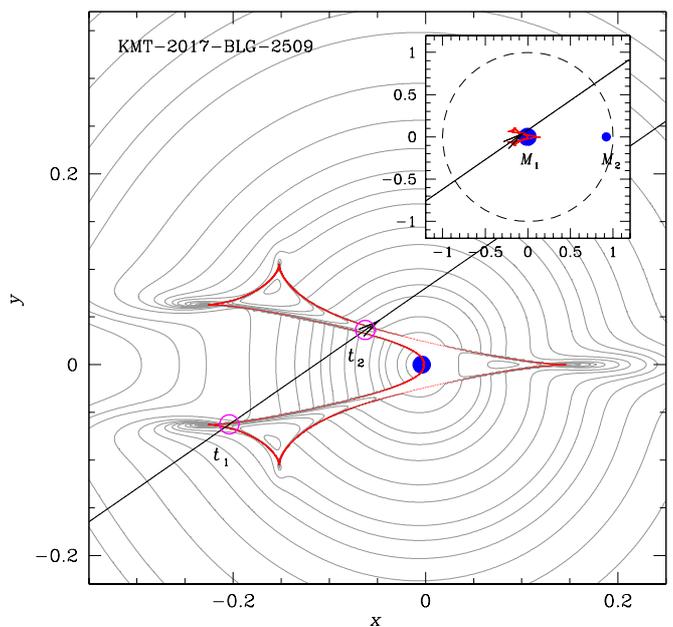}
\caption{
Lens system configuration of KMT-2017-BLG-2509. The inset shows the whole view of the lens system, 
and the main panel shows the enlarged view around the caustic. The red closed figure is the caustic, 
the line with an arrow represents the source motion, the two filled blue dots indicate the positions 
of the lens components, and the dashed circle is the Einstein ring. The grey curves around the caustic 
represent equi-magnification contours. The empty magenta dots on the source trajectory represent the 
source positions at $t_1$ and $t_2$, marked in Fig.~\ref{fig:two}. The size of the dot is not scaled
to the source size.  
}
\label{fig:four}
\end{figure}

\subsection{KMT-2017-BLG-2509}\label{sec:three-one}

Figure~\ref{fig:two} shows the lensing light curve of KMT-2017-BLG-2509. It is characterized by two 
distinctive caustic features that appear at $t_1\sim 7858$ and $t_2\sim 7870$ in ${\rm HJD}^\prime
\equiv {\rm HJD}-2450000$.  For the first caustic feature, both the rising (covered by 4 KMTA data 
points) and falling sides (by a single KMTC point) were captured, while only the rising side was covered 
(by 5 KMTC points) for the second feature. See the zoom-in view of the anomaly region shown in 
Figure~\ref{fig:three}. The region between the two caustic features exhibits negative deviations from a 
single-lens single-source (1L1S) light curve. To be noted about the light curve is that the error bars 
of the data are substantial due to the faintness of the source, and the anomaly region occupies a large 
portion of the magnified region of the light curve. This made it difficult for the anomalies to be 
readily noticed as a planetary (rather than a stellar-binary) signal.\footnote{ The caustic morphology 
and mass ratio of KMT-2017-BLG-2509 is similar to those of MOA-2009-BLG-387 \citep{Batista2011}.  When 
there were relatively few microlensing events being discovered, MOA-2009-BLG-387 was easily identified 
as a planet, while the planetary nature of KMT-2017-BLG-2509 had not been noticed until this paper.  
This indicates that more rigorous review of anomalies in the previous microlensing data is important 
for the accurate estimation of microlensing planet statistics.  }

The 2L1S modeling yields binary parameters of $(s, q)\sim (0.93, 4.4\times 10^{-3})$, indicating 
that the companion to the primary lens is a planetary-mass object lying near the Einstein ring 
of the primary.  We find a unique solution without any degeneracy. For a lensing event produced 
by a binary lens with a small mass ratio, there often exists a pair of close ($s<1.0$) and wide 
($s>1.0$) solutions arising from the similarity between the central caustics induced by the 
companions with $s$ and $s^{-1}$: the close--wide degeneracy \citep{Griest1998, Dominik1999}.  
It is found that KMT-2017-BLG-2509 is not subject to this degeneracy because the light-curve 
morphology (two pairs of caustic crossings that flank a demagnified region) is strictly 
characteristic of an $s<1$ geometry. See Figure~\ref{fig:four}.  The normalized source radius, 
$\rho\sim 1.9 \times 10^{-3}$, is measured by analyzing the caustic-crossing parts of the light 
curve.

The model curve of the 2L1S solution is drawn over the data points in Figures~\ref{fig:two} and 
\ref{fig:three}. In Table~\ref{table:four}, we list the full lensing parameters of the model
together with the flux parameters of the source, $f_s$, and blend, $f_b$, in which the flux is 
approximately scaled to that of $I=18$, that is, $f=10^{-0.4(I-18)}$.
Figure~\ref{fig:four} shows the configuration of the lens system corresponding to the solution.  
In the figure, the line with an arrow indicates the source motion, the red closed figure is the 
caustic, the blue dots, marked by $M_1$ and $M_2$ in the inset, denote the lens positions, and the 
dashed circle represents the Einstein ring.  The planet is located close to the Einstein ring, and 
this results in a single, six-sided resonant caustic. The caustic appears as the merging of the single 
central caustic and the two sets of planetary caustics. The source crossed the tip of the lower 
planetary caustic at $t_1$, and then crossed the slim bridge connecting the upper planetary caustic 
and the central caustic at $t_2$. We mark the source positions corresponding to $t_1$ and $t_2$ by 
empty magenta dots. We note that the size of the dot is not scaled to the source size.  The negative 
deviation region between $t_1$ and $t_2$ occurred when the source passed through the negative deviation
region formed between the two planetary caustics.

\begin{figure}[t]
\includegraphics[width=\columnwidth]{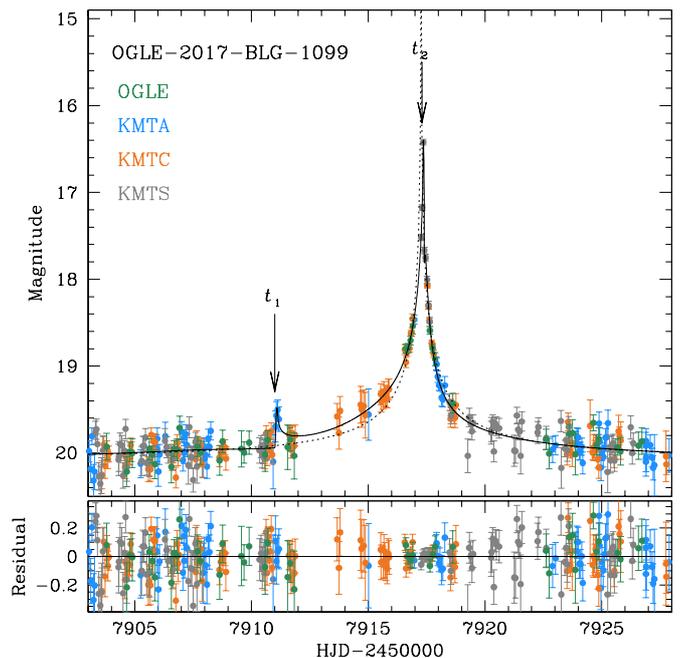}
\caption{
Light curve of OGLE-2017-BLG-1099.  Notations are same as those in Fig.~\ref{fig:two}.
}
\label{fig:five}
\end{figure}

\begin{figure}[t]
\includegraphics[width=\columnwidth]{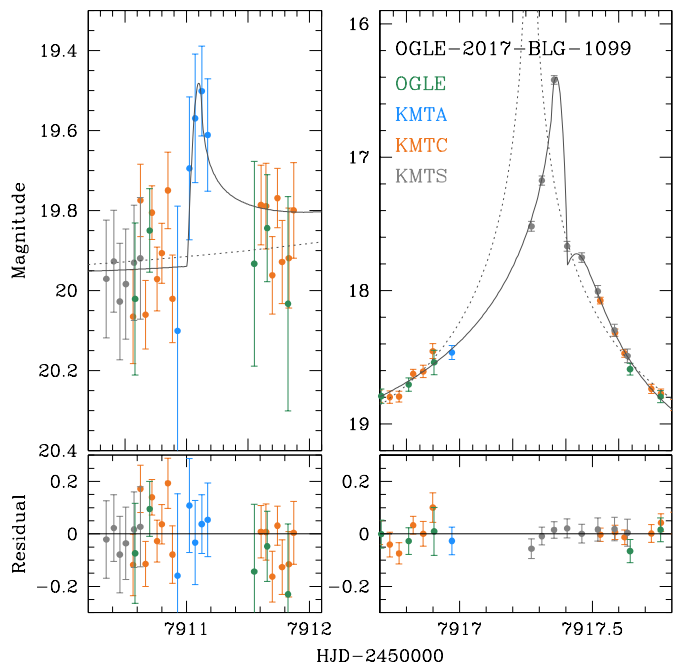}
\caption{
Zoom-in view of the major anomaly region of OGLE-2017-BLG-1099.
}
\label{fig:six}
\end{figure}

\begin{figure}[h]
\includegraphics[width=\columnwidth]{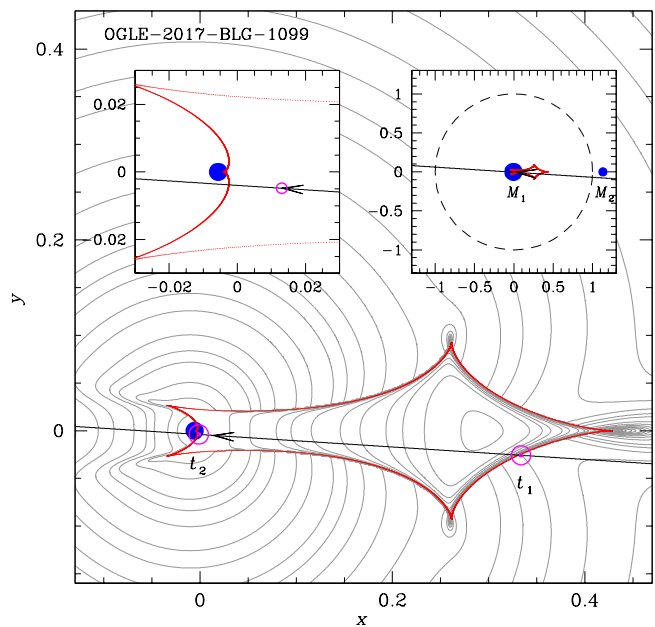}
\caption{
Lens system configuration of OGLE-2017-BLG-1099. Notations are same as those in Fig.~\ref{fig:four}, 
except that an additional inset (left) is presented to show the zoom-in view of the central region.
The source positions corresponding to $t_1$ and $t_2$ designated in Fig.~\ref{fig:five} are marked 
by the empty magenta dots.  The size of the dot in the main panel is arbitrarily set, but the dot 
in the left inset is scaled to the source size.
}
\label{fig:seven}
\end{figure}

\subsection{OGLE-2017-BLG-1099}\label{sec:three-two}

The lensing light curve of OGLE-2017-BLG-1099 is shown in Figure~\ref{fig:five}. The online data 
of the event posted on the alert web pages of the individual survey 
groups\footnote{http://ogle.astrouw.edu.pl/ogle4/ews/ews.html for the OGLE survey, and 
https://kmtnet.kasi.re.kr/~ulens/ for the KMTNet survey.} did not show an obvious anomaly feature 
in the light curve. Nevertheless, the event was selected for a detailed analysis because it reached a 
very high magnification, $A_{\rm max}\sim 140$, during which the chance of detecting a planet-induced 
anomaly was high \citep{Griest1998}.  Optimal light curve obtained by rereducing the data revealed that 
the light curve was anomalous.  The anomaly is characterized first by the asymmetry of the light curve, 
and second by the caustic-involved feature at $t_2\sim 7917.4$.  See the zoom-in view of the region 
around $t_2$ presented in the right panel of Figure~\ref{fig:six}.  It shows that the five KMTS data 
points exhibit the characteristic pattern of magnification variation occurring when a source exits a 
caustic.  See Figure~1 of \citet{Gould1999}.

\begin{figure}[t]
\includegraphics[width=\columnwidth]{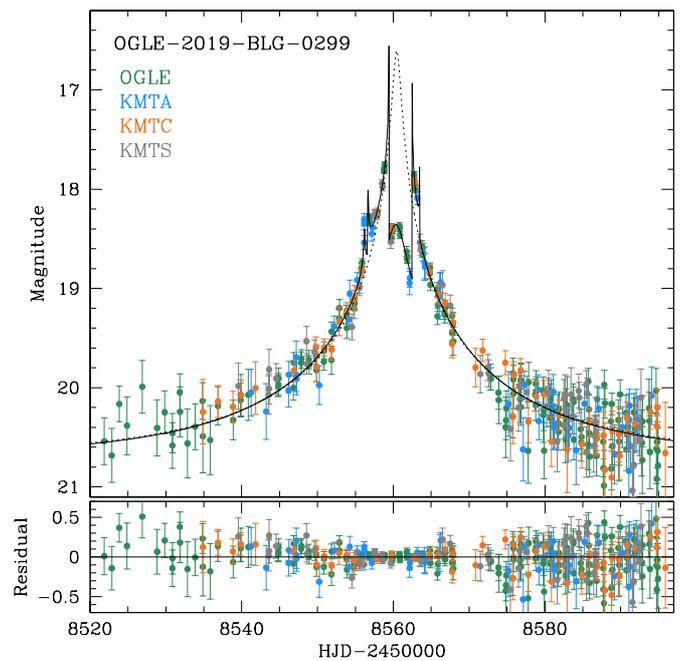}
\caption{
Lensing light curve of OGLE-2019-BLG-0299. Notations are same as those in Fig.~\ref{fig:two}.
}
\label{fig:eight}
\end{figure}

The 2L1S modeling yields a unique solution with the binary parameters of $(s,q)\sim (1.1, 6.4\times 10^{-3})$, 
indicating that the companion is a planetary-mass object with a separation similar to $\thetae$. 
We note that the analysis of the event was done in the 2017 season and its planetary nature 
was realized by one of the coauthors (Y.-H.~Ryu), but the result was not shared with the other 
coauthors.  As a result, the analysis presented in this work  is carried out independently, reaching 
at a result that is consistent with the previous one.  The full lensing parameters are listed in 
Table~\ref{table:four}, and the model curve is presented in Figures~\ref{fig:five} and \ref{fig:six}. 
Due to the proximity of the binary separation to unity, the binary lens pair forms a single resonant 
caustic.  According to the model, the source entered the caustic at $t_1\sim 7911$, and exited the 
caustic at $t_2$.  Due to the weakness of the caustic combined with the low photometric precision, 
the anomaly feature occurred at around the caustic entrance was difficult to be noticed in the preliminary 
modeling using the online data, but the rereduced data showed that the caustic was covered by four KMTA data 
points, although the uncertainties of the data points around the caustic exist were still large.  See the 
zoom-in view around $t_1$ shown in the left panel of Figure~\ref{fig:six}.  The coverage of both the caustic 
entrance and exit yields a normalized source radius of $\rho\sim 1.5\times 10^{-3}$.

Figure~\ref{fig:seven} shows the configuration of the lens system. Like the case of KMT-2017-BLG-2509, 
the caustic appears as the merging of the planetary and central caustics. The source moved approximately 
in parallel with the binary axis ($\alpha\sim 3.7^\circ$).  It entered the caustic by passing the lower 
right side of the planetary caustic, generating a weak caustic spike at $t_1$. Then, the source exited 
the caustic by passing the left side of the central caustic, and then passed the region close to the 
back-end cusp of the central caustic, and this produced the caustic feature at $t_2$.  See the 
zoom-in view of the central region presented in the left inset.  The two empty magenta dots on the 
source trajectory of the main panel represent the source positions at $t_1$ and $t_2$.  We note that 
the size of these dots is arbitrarily set, but the source dot in the left inset is scaled to the source 
size.

\begin{figure}[t]
\includegraphics[width=\columnwidth]{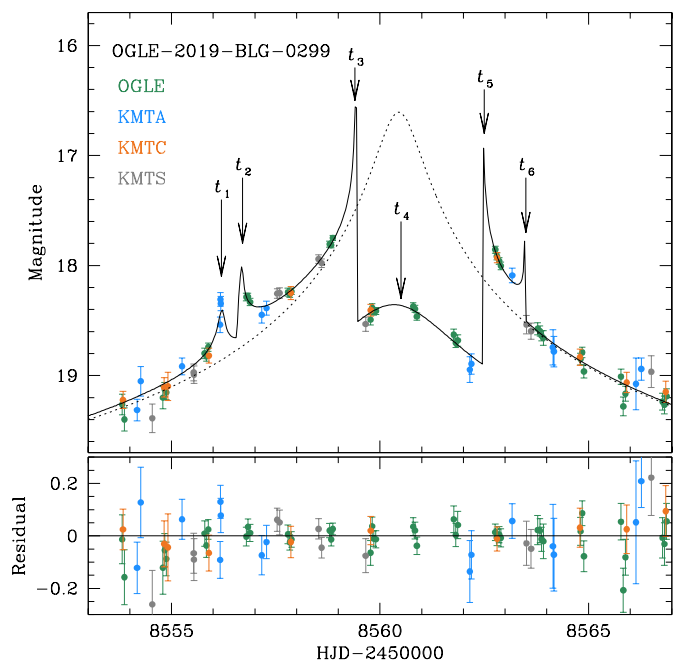}
\caption{
Enlarged view around the anomaly region of OGLE-2019-BLG-0299 light curve. The positions marked by 
arrows and labeled as $t_1$, $t_2$, $t_3$, $t_4$, $t_5$, and $t_6$, represent the epochs of major 
anomaly features.
}
\label{fig:nine}
\end{figure}

\subsection{OGLE-2019-BLG-0299}\label{sec:three-three}

Figure~\ref{fig:eight} shows the light curve of OGLE-2019-BLG-0299.  The anomalous nature of the 
light curve was known when the event was proceeding, and planetary-lensing models found with the 
use of the OGLE data were circulated to the microlensing community by C.~Han and V.~Bozza before 
the end of the event.  The analysis in this work is done with the addition of the KMTNet data 
obtained from the optimized photometry of the event.  The enlarged view of the light curve around 
the anomaly region is shown in Figure~\ref{fig:nine}.  The anomaly exhibits a very complex pattern, 
that is characterized by 6 peaks or bumps at $t_1\sim 8556.2$, $t_2\sim 8556.7$, $t_3\sim 8559.4$, 
$t_4\sim 8560.5$, $t_5\sim 8562.5$, and $t_6\sim 8563.5$.

A 2L1S modeling yields a unique solution with the binary parameters of $(s, q)\sim (0.99, 10.0\times 
10^{-3})$.  The binary separation is very close to unity, as in the cases of the two previous events.  
The mass ratio is about ten times the Jupiter/sun ratio of the solar system, but the mass of the 
companion is still in the planet-mass regime, considering that the measured event time scale, $\te \sim 
30$~days, is not much longer than those of typical lensing events produced by low-mass stars. The full 
lensing parameters are presented in Table~\ref{table:four}. Although all the major anomaly features were 
delineated, none of the caustics was resolved densely enough for the secure measurement of $\rho$, and 
only the upper limit, $\rho_{\rm max}\sim 3.5\times 10^{-3}$, is constrained.

\begin{figure}[t]
\includegraphics[width=\columnwidth]{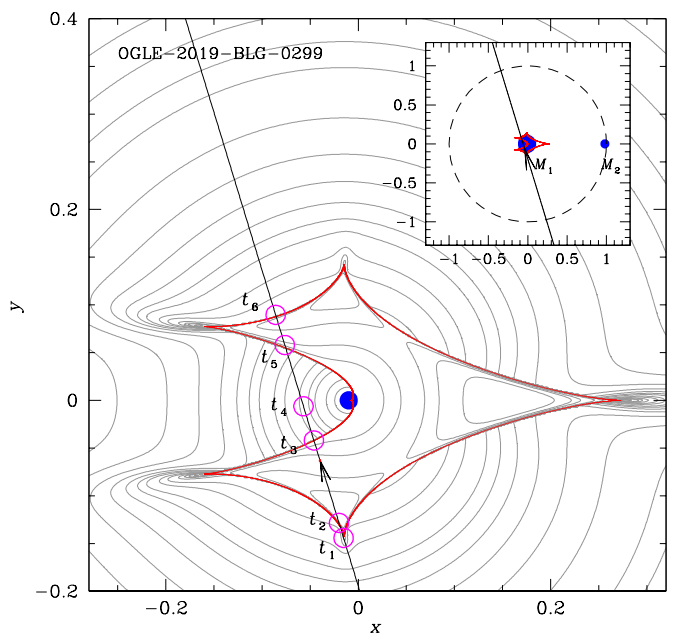}
\caption{
Lens system configuration of OGLE-2019-BLG-0299. The empty magenta points represent the source 
positions corresponding to the six epochs of $t_1$, $t_2$, $t_3$, $t_4$, $t_5$, and $t_6$, that 
are marked by arrows in Fig.~\ref{fig:nine}.  The size of the dots is not scaled.
}
\label{fig:ten}
\end{figure}

Figure~\ref{fig:ten} shows the lens system  configuration of OGLE-2019-BLG-0299.  According to the 
model, the source crossed the binary axis with a source trajectory angle of $\alpha\sim 73^\circ$, 
and passed through the 6-fold resonant caustic four times at $t_2$, $t_3$, $t_5$, and $t_6$, where 
$(t_2, t_3)$ and $(t_5, t_6$) are the two time pairs of caustic entrance and exit. These caustic 
crossings produced the spikes at the corresponding times, and the regions of the light curve between 
the individual caustic-crossing pairs exhibited characteristic U-shape trough patterns. The bump at 
$t_1$ was produced by the source approach close to the lower cusp of the caustic before the first 
caustic crossing at $t_2$. The other bump at $t_4$ was produced when the source passed through the 
outer-caustic region near the left-side on-axis cusp of the caustic. The source positions corresponding 
to the 6 epochs of the anomaly are marked by empty magenta dots, of which size is not scaled, on the 
source trajectory.

\begin{table*}[t]
\small
\caption{Source color, magnitude, radius, Einstein radius, and proper motion\label{table:five}}
\begin{tabular}{llllcc}
\hline\hline
\multicolumn{1}{c}{Quantity}                &
\multicolumn{1}{c}{KMT-2017-BLG-2509}       &
\multicolumn{1}{c}{OGLE-2017-BLG-1099}      &
\multicolumn{1}{c}{OGLE-2019-BLG-0299}      \\
\hline
$(V-I, I)$                       &  $(3.482\pm 0.186, 23.013\pm 0.077)$   &   $(3.751\pm 0.172, 22.795\pm 0.069)$    &   $(2.488\pm 0.126, 20.986\pm 0.039)$  \\
$(V-I, I)_{\rm RGC}$             &  $(3.401, 17.277)                  $   &   $(3.785, 17.327)                  $    &   $(2.653, 15.974)                  $  \\
$(V-I, I)_{{\rm RGC},0}$         &  $(1.060, 14.441)                  $   &   $(1.060, 14.445)                  $    &   $(1.060, 14.439)                  $  \\
$(V-I, I)_0$                     &  $(1.141\pm 0.186, 20.177\pm 0.077)$   &   $(1.026\pm 0.172, 19.912\pm 0.069)$    &   $(0.894\pm 0.126, 19.451\pm 0.039)$  \\
$\theta_*$ ($\mu$as)             &  $0.47 \pm 0.09                    $   &   $0.47 \pm  0.09                   $    &   $0.50 \pm  0.07                   $  \\
$\thetae$ (mas)                  &  $0.243 \pm 0.110                  $   &   $0.304 \pm 0.125                  $    &   $> 0.14                           $  \\
$\mu$ (mas/yr)                   &  $1.32 \pm 0.60                    $   &   $5.89 \pm 2.47                    $    &   $> 1.74                           $  \\
\hline
\end{tabular}
\end{table*}

\section{Source stars and Einstein radii}\label{sec:four}

In this section, we estimate the angular Einstein radii of the events. The Einstein 
radius is determined by $\thetae =\theta_*/\rho$. The $\rho$ value is measured or constrained 
from the analysis of the caustic-crossing parts, which are affected by finite-source effects. 
With the measured $\rho$, then it is required to estimate the angular source radius $\theta_*$. 
The value of $\theta_*$ is deduced from the color and magnitude of the source.

In general, the source color is estimated by measuring the source
magnitudes in two passbands, $I$ and $V$ bands in our case, by regressing the data of the 
individual passbands with the variation of the lensing magnification. The $I$-band magnitudes 
are securely measured by this method for all events.  However, the $V$-band magnitudes are 
difficult  to be measured by this method due to the large uncertainties of the data, making 
it difficult to securely estimate the source colors.  We, therefore, estimate the source color 
using the \citet{Bennett2008} method, which utilizes the CMD constructed from the {\it Hubble 
Space Telescope} ({\it HST}) observations \citep{Holtzman1998}. In the first step of this method, 
the {\it HST} CMD is aligned with the CMD obtained from ground-based observations using the 
well defined centroid of the red giant clump (RGC). In the second step, the source position 
in the CMD is interpolated from the branch of main-sequence or giant stars on the {\it HST} 
CMD using the well measured $I$-band magnitude difference between the RGC centroid and the 
source. In the final step, the source color and its uncertainty are estimated as the mean and 
standard deviation of stars located on the branch.

\begin{figure}[t]
\includegraphics[width=\columnwidth]{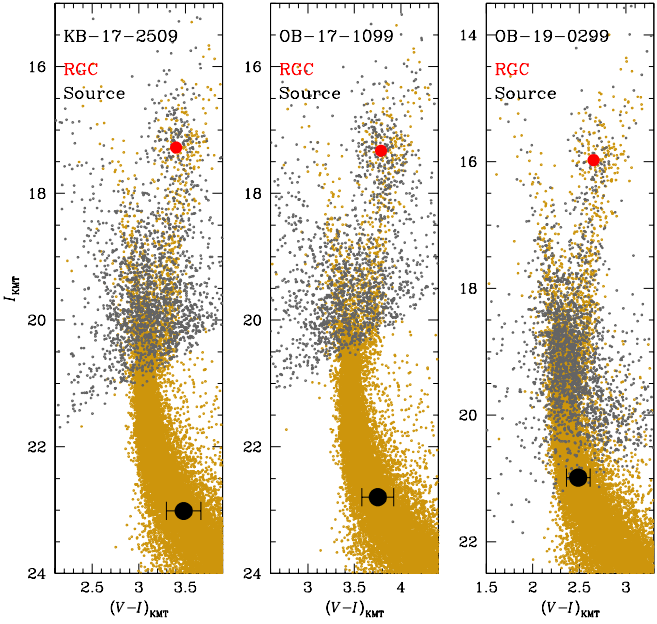}
\caption{
Source positions (filled black dot with error bars) in the instrumental CMD with respect to the 
centroids of red giant clump (RGC, filled red dots) for the individual lensing events. In each 
panel, the ground-based and HST CMDs are marked in grey and brown dots, respectively.  
}
\label{fig:eleven}
\end{figure}

Figure~\ref{fig:eleven} shows the source locations (black filled dots with error bars) with 
respect to the RGC centroids (red filled dot) in the combined CMD, in which the {\it HST} 
CMD is represented by brown dots and the CMD from the ground-based observations is represented 
by grey dots.  In Table~\ref{table:five}, we list the positions of the source, $(V-I, I)$, and 
the RGC centroid, $(V-I, I)_{\rm RGC}$, measured on the instrumental CMD. Then, the reddening 
and extinction corrected (de-reddened) color and magnitude of the source, $(V-I, I)_0$, are 
determined using the offsets from the RGC centroid, $\Delta (V-I, I)$, together with the known 
de-reddened values of RGC, $(V-I, I)_{{\rm RGC},0}$ \citep{Bensby2013, Nataf2013}, by the relation
\begin{equation}
(V-I, I)_0 = (V-I, I)_{{\rm RGC},0} + \Delta (V-I, I).
\label{eq1}
\end{equation}
The values of $(V-I, I)_0$, $\Delta (V-I, I)$, $(V-I, I)_{{\rm RGC},0}$ for the individual 
events are listed in Table~\ref{table:five}.  We note that the de-reddened $I$-band magnitudes 
of the RGC centroids, that is, $I_{{\rm RGC},0}$, vary depending on the event, because we 
consider the varying distance depending on the source location using Table~1 of \citet{Nataf2013}.  
The measured colors and magnitudes, which are in the ranges of $0.9\lesssim (V-I)_0\lesssim 1.1$ 
and $19.5\lesssim I_0\lesssim 20.2$, respectively, indicate that the source stars of the events
have similar spectral types of early K-type main sequence stars.

The measured $(V-I)$ color is then converted into $(V-K)$ color using the color-color 
relation of \citet{Bessell1988}, and then the source radius is deduced from the $(V-K)$--$\theta_*$ 
relation of \citet{Kervella2004}. With the measured source radius, then, the Einstein radius and 
the relative lens-source proper motion are estimated by the relations $\thetae = \theta_*/\rho$ 
and $\mu  = \thetae/\te$, respectively.  We list the estimated values of $\theta_*$, $\thetae$, 
and $\mu$ in Table~\ref{table:five}.  We note that the lower limits $\thetae$ and $\mu$ are 
presented for OGLE-2019-BLG-0299, because only the upper limit of $\rho$ is constrained for the 
event.

\begin{table*}[t]
\small
\caption{Physical lens parameters\label{table:six}}
\begin{tabular}{llllcc}
\hline\hline
\multicolumn{1}{c}{Parameter}               &
\multicolumn{1}{c}{KMT-2017-BLG-2509}       &
\multicolumn{1}{c}{OGLE-2017-BLG-1099}      &
\multicolumn{1}{c}{OGLE-2019-BLG-0299}      \\
\hline
$M_{\rm planet}$ ($M_{\rm J}$)  &  $2.09^{+1.68}_{-1.26}$   &   $3.02^{+2.43}_{-1.81}$    &  $6.22^{+3.80}_{-3.67}$  \\
$M_{\rm host}$ ($M_\odot$)      &  $0.46^{+0.37}_{-0.27}$   &   $0.45^{+0.36}_{-0.27}$    &  $0.59^{+0.36}_{-0.35}$  \\
$\dl$ (kpc)                     &  $7.04^{+0.89}_{-1.27}$   &   $7.25^{+1.06}_{-1.40}$    &  $5.83^{+1.21}_{-1.85}$  \\
$a_\perp$ (AU)                  &  $2.14^{+0.27}_{-0.39}$   &   $2.73^{+0.40}_{-0.53}$    &  $2.80^{+0.58}_{-0.89}$  \\
\hline
\end{tabular}
\end{table*}

\section{Physical lens parameters}\label{sec:five}

We estimate the physical lens parameters of the lens mass, $M$, and distance, $\dl$, by 
conducting a Bayesian analysis. The lensing observables that can be used to constrain $M$ 
and $D_{\rm L}$ include $\te$, $\thetae$, and $\pie$, where the first two observables are 
related to $M$ and $\dl$ by
\begin{equation}
\te = {\thetae\over \mu};\ \ \thetae=(\kappa M \pi_{\rm rel})^{1/2};\ \ 
\pi_{\rm rel}={\rm AU}\left( {1\over D_{\rm L}}-{1\over D_{\rm S}} \right),
\label{eq2}
\end{equation}
and the additional measurement of $\pie$ would allow one to uniquely determine the lens 
parameters by
\begin{equation}
M = {\thetae \over \kappa\pie };\qquad \dl = {{\rm AU} \over \pie\thetae + \pi_{\rm S} }.
\label{eq3}
\end{equation}
Here $\kappa=4G/(c^2{\rm AU})$, $\pi_{\rm S}={\rm AU}/D_{\rm S}$, and $D_{\rm S}$ denotes 
the distance to the source \citep{Gould2000}.  The available observables vary depending on 
the events: $\te$ and $\thetae$ for KMT-2017-BLG-2509 and OGLE-2017-BLG-1099, and $\te$ 
and the lower limit of $\thetae$ for OGLE-2019-BLG-0299.  The value of $\pie$ is not measured 
for any of the events.

In the first step of the Bayesian analysis, we produce artificial lensing events by conducting 
a Monte Carlo simulation. In the simulation, we use a prior Galactic model, which describes the 
physical and dynamical distributions and the mass function of Galactic objects. We adopt the 
\citet{Jung2021} Galactic model, in which the physical distribution of disk and bulge objects 
are described by the \citet{Robin2003} and \citet{Han2003} models, respectively, the dynamical 
distributions of the disk and bulge objects are depicted by the \citet{Jung2021} and \citet{Han1995} 
models, respectively, and the \citet{Jung2018} mass function model is commonly used for both 
populations. In the second step, we choose events with $\te$ and $\thetae$ values consistent 
with the measured observables, and construct the posterior distributions of $M$ and $\dl$ for 
these events. Then, the representative values of the lens parameters are determined as the 
median values of the posterior distributions, and the lower and upper limits are determined 
as 16\% and 84\% of the distributions, respectively.

\begin{figure}[t]
\includegraphics[width=\columnwidth]{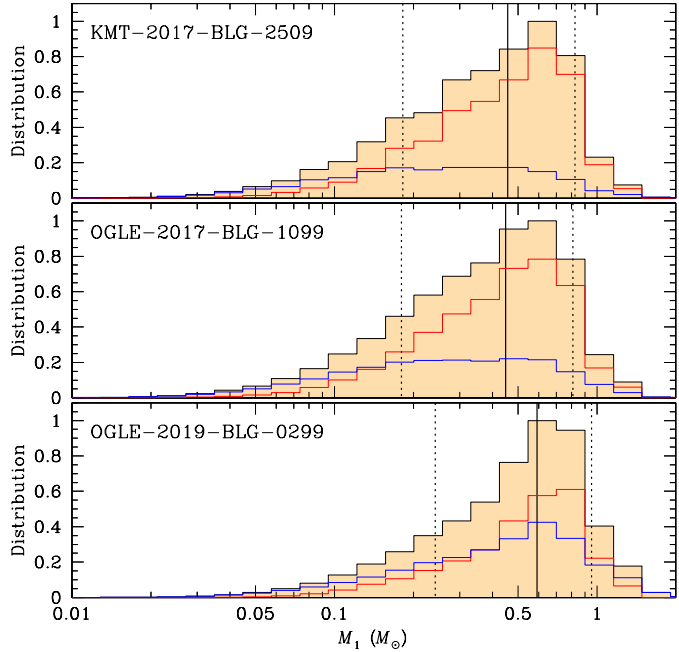}
\caption{
Bayesian posteriors of the primary lens mass, $M_1$, for the KMT-2017-BLG-2509 (top panel), 
OGLE-2017-BLG-1099 (middle panel), and OGLE-2019-BLG-0299 (bottom panel). In each panel, the 
blue and red curves represent the contributions by the disk and bulge lens populations, 
respectively, and the black curve represents the sum of the two lens populations.  The solid 
vertical line represents the median, and the two dotted lines indicate the 1$\sigma$ range of 
the distribution.
}
\label{fig:twelve}
\end{figure}

\begin{figure}[t]
\includegraphics[width=\columnwidth]{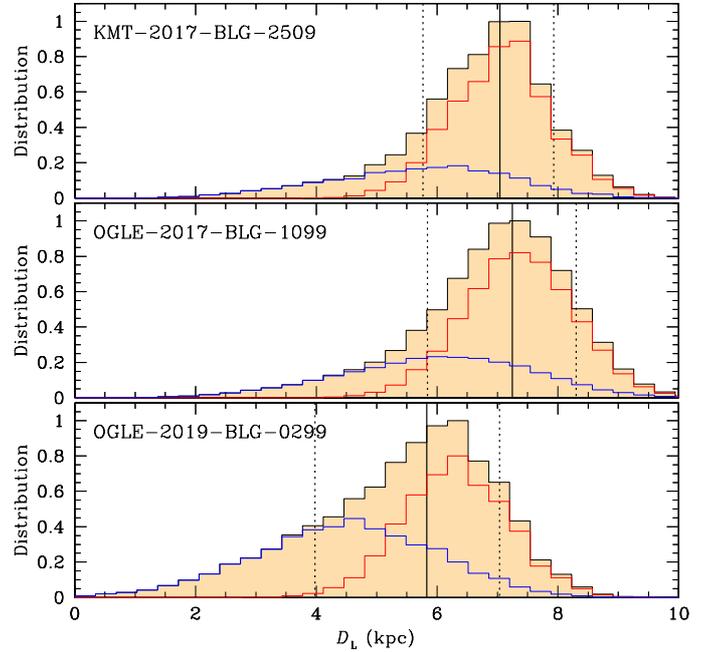}
\caption{
Bayesian posteriors of the lens distance, $\dl$. Notations are same as those in Fig.~\ref{fig:twelve}.
}
\label{fig:thirteen}
\end{figure}

In Figures~\ref{fig:twelve} and \ref{fig:thirteen} , we present the Bayesian posterior distributions 
of $M_1$ and $\dl$, respectively. In Table~\ref{table:six}, we list the estimated masses of the 
host, $M_{\rm host}\equiv M_1$, and planet, $M_{\rm planet}\equiv M_2$, distance, and projected 
planet-host separation, $a_\perp$. The projected separation is calculated by $a_\perp=s\dl \thetae$. 
It is estimated that the host masses are in the range of $0.45\lesssim M_{\rm host}/M_\odot\lesssim 
0.59$, which corresponds to early M to late K dwarfs, and thus the host stars are less massive than 
the sun.  On the other hand, the planet masses, which are in the range of $2.1\lesssim M_{\rm planet}
/M_{\rm J}\lesssim 6.2$, are heavier than the mass of the heaviest planet of the solar system, that is, 
Jupiter.  Considering that the snow line distance is $a_{\rm sl}\sim 2.7(M/M_\odot)$~AU, and $a_\perp$ 
is the separation in projection, the planets in all systems lie beyond the snow lines of the hosts. 
Therefore, the discovered planetary systems, together with many other microlensing planetary systems, 
support that  massive gas-giant planets are commonplace around low-mass stars.  See the distribution 
of exoplanets with respect to $a_\perp/a_{\rm sl}$ presented in Figure~6 of \citet{Gaudi2012}.  
The contributions to the posterior distribution by the disk and bulge lens populations are marked 
by blue and red curves, respectively.  The disk/bulge contributions are 28\%/72\% for KMT-2017-BLG-2509, 
33\%/67\% for OGLE-2017-BLG-1099, and 48\%/52\% for OGLE-2019-BLG-0299.  The relatively low disk 
contribution for KMT-2017-BLG-2509 arises due to the constraint of the low relative lens-source proper 
motion, $\mu\sim 1.3$~mas/yr, because low proper motion is difficult to produce for disk lenses.

\section{High-resolution follow-up observation}\label{sec:six}

In the general case of a lensing event, high-resolution observations using space-borne telescopes 
or ground-based adaptive optics instruments allow one to measure the lens flux and the lens-source 
separation.  The lens flux allows one to estimate the lens mass $M$, and the lens-source separation 
allows one to estimate the relative lens-source proper motion $\mu$, which in turn constrains the 
Einstein radius by $\thetae=\mu\te$.

For faint-source planetary lensing events with resonant caustic features, high-resolution follow-up 
observations are especially important for clarifying the planetary lens systems.  The size of a 
resonant planetary caustic scales to the planet/host mass ratio as $\Delta\zeta_{\rm c}\propto q^{1/3}
\thetae$, and thus the duration of the planetary anomaly is related to the event time scale by 
$\Delta t = \Delta \zeta_{\rm c}/\mu \propto q^{1/3}\te$.  For a given anomaly duration, then, the 
event time scale and the mass ratio are related by $q \propto t_{\rm E}^{-3}$.  For faint-source 
events, the very faint sources can make it difficult to precisely determine $\te$, and this leads 
to a large uncertainty of $q$, because $\Delta q \propto 3 \Delta \te$.  Late time observations in 
two passbands can yield the source flux and color, and so $I$-band source-flux estimate.  The well 
estimated source flux can further constrain $\te$, and this leads to a tight constraint on the 
planet mass ratio.  High-resolution image data could have resulted in best constraints if they had 
been acquired at the time of the event to provide a comparison.  Unfortunately, no such images were 
taken because the planetary nature of the events were not known at the times of the event discoveries.  
However, post-event imaging can still help to constrain the physical parameters the lens systems.

\section{Summary and conclusion}\label{sec:seven}

We reported three microlensing planets KMT-2017-BLG-2509Lb, OGLE-2017-BLG-1099Lb, and 
OGLE-2019-BLG-0299Lb that were found from the reinvestigation of the microlensing data collected 
by the KMTNet and OGLE surveys during the 2017--2019 seasons.  For all of these lensing events, 
the planetary signals deviated from the typical form of short-term anomalies, because they were 
produced by the crossings of the source stars over the resonant caustics formed by the giant planets 
located at around the Einstein rings of host stars.  The faintness of the source stars  and the 
relatively low observational cadences also contributed to the difficulty of finding the planetary 
signals.  Due to the resonant nature of the caustics, the lensing solutions were uniquely determined 
without any degeneracy.  The estimated masses of the planet hosts are in the range of 
$0.45\lesssim M/M_\odot \lesssim 0.59$, which corresponds to early M to late K dwarfs, and thus 
the host stars are less massive than the sun.  On the other hand, the planets, with masses in the 
range of $2.1\lesssim M_{\rm planet}/M_{\rm J}\lesssim 6.2$, are heavier than Jupiter of the solar 
system.  The planets in all systems lie beyond the snow lines of the hosts, and thus the discovered 
planetary systems support the conclusion that massive gas-giant planets are commonplace around 
low-mass stars.

\begin{acknowledgements}
Work by C.H. was supported by the grants of National Research Foundation of Korea
(2019R1A2C2085965 and 2020R1A4A2002885).
This work was conducted during the research year of Chungbuk National University in 2021.
This research has made use of the KMTNet system operated by the Korea
Astronomy and Space Science Institute (KASI) and the data were obtained at
three host sites of CTIO in Chile, SAAO in South Africa, and SSO in
Australia.
The OGLE project has received funding from the National Science Centre, Poland, grant
MAESTRO 2014/14/A/ST9/00121 to AU.
\end{acknowledgements}


\begin{thebibliography}{}

\bibitem[Alard \& Lupton(1998)]{Alard1998} Alard, C., \& Lupton, R.~H.\ 1998, \apj, 503, 325
\bibitem[Albrow(2017)]{Albrow2017} Albrow, M.\ 2017, MichaelDAlbrow/pyDIA: Initial Release on Github,Version v1.0.0, Zenodo, doi:10.5281/zenodo.268049
\bibitem[Albrow et al.(2009)]{Albrow2009} Albrow, M., Horne, K., Bramich, D.~M., et al.\ 2009, \mnras, 397, 2099
\bibitem[Alcock et al.(1993)]{Alcock1993} Alcock, C., Akerlof, C.~W., Allsman, R.~A., et al.\ 1993, \nat, 365, 621
\bibitem[An \& Han(2002)]{An2002} An, J.~H., \& Han, C.\ 2002, \apj, 573, 351
\bibitem[Aubourg et al.(1993)]{Aubourg1993} Aubourg, E., Bareyre, P., Br\'ehin, S., et al.\ 1993, \nat, 365, 623
\bibitem[Batista et al.(2011)]{Batista2011} Batista, V., Gould, A., Dieters, S., et al.\ 2011, \aap, 529, A102
\bibitem[Bennett et al.(2008)]{Bennett2008} Bennett, D.~P., Bond, I.~A., Udalski, A., et al. 2008, \apj, 684, 663B
\bibitem[Bensby et al.(2013)]{Bensby2013} Bensby, T., Yee, J. C., Feltzing, S., et al.\ 2013, \aap, 549, 14
\bibitem[Bessell \& Brett(1988)]{Bessell1988} Bessell, M.~S., \& Brett, J.~M.\ 1988, \pasp, 100, 1134
\bibitem[Claret(2000)]{Claret2000} Claret, A.\ 2000, \aap, 363, 1081
\bibitem[Di Stefano \& Mao(1996)]{Stefano1996} Di Stefano, R., \& Mao, S.\ 1996, \apj, 457, 93
\bibitem[Dominik(1998)]{Dominik1998} Dominik, M.\ 1998, \aap, 329, 361
\bibitem[Dominik(1999)]{Dominik1999} Dominik, M.\ 1999, \aap, 349, 108
\bibitem[Gaudi(2012)]{Gaudi2012} Gaudi, B.~S. 2012, \araa, 50, 411
\bibitem[Gould(1992)]{Gould1992} Gould, A.\ 1992, \apj, 392, 442
\bibitem[Gould(2000)]{Gould2000} Gould, A.\ 2000, \apj, 542, 785
\bibitem[Gould \& Andronov(1999)]{Gould1999} Gould, A., \& Andronov, N.\ 1999, \apj, 516, 236
\bibitem[Griest \& Safizadeh(1998)]{Griest1998} Griest, K., \& Safizadeh, N.\ 1998, \apj, 500, 37
\bibitem[Han \& Gould(1995)]{Han1995} Han, C., \& Gould, A.\ 1995, \apj, 447, 53
\bibitem[Han \& Gould(2003)]{Han2003} Han, C., \& Gould, A.\ 2003, \apj, 592, 172
\bibitem[Han et al.(2020a)]{Han2020a} Han, C., Kim, D., Udalski, A., et al.\ 2020a, \aj, 160, 64  
\bibitem[Han et al.(2020b)]{Han2020b} Han, C., Shin, I.-G., Jung, Y.~K., et al.\ 2020b, \aap, 641, A105
\bibitem[Han et al.(2021a)]{Han2021a} Han, C., Udalski, A., Kim, D., et al. 2021a, \aap, 642, 110
\bibitem[Han et al.(2021b)]{Han2021b} Han, C., Udalski, A., Kim, D., et al. 2021b, \aap, 650, A89
\bibitem[Han et al.(2021c)]{Han2021c} Han, C., Udalski, A., Kim, D., et al. 2021c, \aap, 649, A90
\bibitem[Holtzman et al.(1998)]{Holtzman1998} Holtzman, J.~A., Watson, A.~M., Baum, W.~A., et al. 1998, \aj, 115, 1946
\bibitem[Hwang et al.(2021)]{Hwang2021} Hwang, W., Zang, W., Gould, A., et al. 2021, in preparation
\bibitem[Jung et al.(2021)]{Jung2021} Jung, Y.~K., Han, C., Udalski, A., et al.\ 2021, \aj, 161, 293
\bibitem[Jung et al.(2018)]{Jung2018} Jung, Y.~K., Udalski, A., Gould, A., et al.\ 2018, \aj, 155, 219
\bibitem[Kervella et al.(2004)]{Kervella2004} Kervella, P., Th\'evenin, F., Di Folco, E., \& S\'egransan, D. 2004, \aap, 426,29
\bibitem[Kim et al.(2016)]{Kim2016} Kim, S.-L., Lee, C.-U., Park, B.-G., et al.\ 2016, JKAS, 49, 37
\bibitem[Mr\'oz et al.(2017)]{Mroz2017} Mr\'oz, P., Han, C., Udalski, A., et al.\ 2017, \aj, 153, 143
\bibitem[Nataf et al.(2013)]{Nataf2013} Nataf, D.~M., Gould, A., Fouqu\'e, P., et al.\ 2013, \apj, 769, 88
\bibitem[Robin et al.(2003)]{Robin2003} Robin, A.~C., Reyl\'e, C., Derri\'ere, S., \& Picaud, S.\ 2003, \aap, 409, 523
\bibitem[Sumi et al.(2013)]{Sumi2013} Sumi, T., Bennett, D.~P., Bond, I.~A., et al.\ 2013, \apj, 778, 150
\bibitem[Tomaney \& Crotts(1996)]{Tomaney1996} Tomaney, A.~B., \& Crotts, A.~P.~S.\ 1996, \aj, 112, 2872
\bibitem[Udalski et al.(1993)]{Udalski1993} Udalski, A., Szyma\'nski, M., Ka{\l}u\.zny, J., et al.\ 1993, \actaa, 43, 289
\bibitem[Udalski et al.(1994)]{Udalski1994} Udalski, A., Szyma\'nski, M., Ka{\l}u\.zny, J., Kubiak, M. Mateo, M., \& Krzemi\'nski, W.\ 1994, \actaa, 44, 1
\bibitem[Udalski et al.(2015)]{Udalski2015} Udalski, A., Szyma\'nski, M.~K., \& Szyma\'nski, G.\ 2015, \actaa, 65, 1
\bibitem[Wo\'zniak(2000)]{Wozniak2000} Wo\'zniak, P. R. 2000, \actaa, 50, 42
\bibitem[Yee et al.(2012)]{Yee2012} Yee, J. C., Shvartzvald, Y., Gal-Yam, A., et al.\ 2012, \apj, 755, 102
\bibitem[Zang et al.(2021)]{Zang2021} Zang, W., Hwang, K.-H., Udalsk, A., et al.\ 2021, \aj, submitted (arXiv:2103.11880)




 \end{thebibliography}
 \end{document}